\documentclass[prd,aps,showpacs,preprintnumbers,amssymb]{revtex4}
\usepackage{axodraw}
\usepackage{color}
\usepackage{epsf}

\def\e3p{$\eta \rightarrow 3 \pi$}

\begin{document}
\title{%
\hfill{\normalsize\vbox{%
\hbox{}
 }}\\
{Note about Yang Mills, QCD and their supersymmetric counterparts}}

\author{Renata Jora $^{\it \bf a}$~\footnote[2]{Email:
 rjora@ifae.es}}
 \affiliation{$^ {\bf \it a}$ Grup de Fisica Teorica and IFAE, Universitat Autonoma de Barcelona,
 E-08193 Bellaterra(Barcelona), Spain.}

\date{\today}

\begin{abstract}
We analyze, in an effective Lagrangian framework, the connection between Super QCD (Super Yang Mills) and QCD
 (Yang Mills) by highlighting the crucial role that the zero modes play in the process of decoupling gluinos and squarks.
\end{abstract}


\maketitle
\section{Introduction}

The effective Lagrangian approach to supersymmetric gauge theories \cite{Veneziano1}- \cite {Seiberg2} can lead to
 valuable insights for the ordinary gauge theories like QCD and Yang Mills.
In the present work we analyze the mechanism which relates Super QCD (Super Yang Mills) with QCD (Yang Mills)
by separating the zero and positive modes contributions in the beta function for all the fields involved.

The all orders beta function for Super QCD (Super Yang-Mills) was established long time ago
using both a perturbative and an instanton approach \cite{Novikov}:
\begin{eqnarray}
\beta(\alpha)&=&
\frac{-\alpha^2}{2\pi}\left[(n_g-\frac{1}{2}n_f)+\frac{1}{2}\sum_{{\rm zero\, modes}}\gamma_{(\psi)}\right]
\nonumber\\
&\times&[1-\frac{\alpha}{4\pi}(n_g-n_{\lambda})]^{-1}
\label{no78}
\end{eqnarray}

Here $n_g=4N_c$ are the gluon zero modes, $n_f=2N_c+2N_f$ counts the zero fermion modes (gluino and quarks), $\gamma_{(\Psi)}$ is the anomalous dimension of the quark mass operator and $n_{\lambda}=2N_c$ is the number of gluino zero modes.

In terms of the number of colors, $N_c$ and flavors, $N_f$ the beta function reads:
\begin{eqnarray}
\beta(g)&=&-\frac{g^3}{16\pi^2}\frac{\beta_0+N_f\gamma(g^2)}{1-\frac{g^2}{8\pi^2}N_c}
\nonumber\\
\gamma(g^2)&=&-\frac{g^2}{4\pi^2}\frac{N^2-1}{2N}
\label{bet5467}
\end{eqnarray}

In first order this beta function can be easily compared to that of QCD; zero modes contributions for gluon and fermion are the same. QCD beta function however does not contain any modes for gluino and squarks but includes the extra contribution from gluon and fermion positive modes. Recently Sannino and Pica \cite {SP} proposed an all order beta function for ordinary non-abelian gauge theories which can also be obtained by simply adding positive modes contributions to the supersymmetric beta function.

We are interested mainly in what happens to the trace of energy momentum tensor when one adds mass terms to the super QCD
Lagrangian. We will show that the decoupling of the heavy degrees of freedom (gluinos and squarks for the cases of interest here) take place in two stages corresponding to the magnitude of the mass term. In the first stage the gluino and squarks zero modes decouple whereas their positive modes are still canceled by those of gluons and quarks, respectively. It is like somehow supersymmetry is still present at the quantum level. For the second stage an imbalance in the positive modes appears which leads to the complete decoupling. In the process we get as a bonus a straightforward way to obtain a value for the gluon condensate from a QCD potential derived from supersymmetry.

In section II we will discuss the connection between super Yang Mills and Yang Mills whereas in section III we will relate
 super QCD and QCD.  The final section is dedicated to conclusions.

\section{Super Yang Mills}

The effective Lagrangian for super Yang Mills theory was derived by Veneziano and Yankielowicz \cite{Veneziano1}
and has the form:
\begin{eqnarray}
{\cal L}=
&&\frac{9}{\alpha}\int d^2\theta d^2{\bar\theta} (S{\bar S})^{\frac{1}{3}}+
\nonumber\\
&&\int d^2\theta S[\ln(\frac{S}{\Lambda^3})^{N_c}-N_c]+ h.c.
\label{langr}
\end{eqnarray}

Here $\Lambda$ is the super Yang Mills invariant scale and $S=\frac{g^2}{32\pi^2} W^{\alpha}_aW_{\alpha,a}$ or more explicitly, $S=\Phi(y)+\sqrt{2}\theta\Psi(y)+\theta^2F(y)$. The super space variable y is given by $y^m=x^m+i\theta \sigma^m {\bar \theta}$ and,

\begin{eqnarray}
\Phi&=&-\frac{g^2}{32\pi^2}\lambda^2
\nonumber\\
\sqrt{2}\Psi&=&\frac{g^2}{32\pi^2}[\sigma^{mn}\lambda_aF_{mn,a}-i\lambda_aD_a]
\nonumber\\
F &=&\frac{g^2}{32\pi^2}[-\frac{1}{2}F^{mn}_aF_{mn,a}-\frac{i}{4}\epsilon_{mnrs}F^{mn}_aF^{rs}_a
\nonumber\\
&+&D_aD_a-i{\bar \lambda}_a{\bar \sigma}^m \vec{\nabla}\lambda_a+i\partial^m J^5_m].
\label{comp87}
\end{eqnarray}

This Lagrangian is built especially to satisfy the three anomalies,
\begin{eqnarray}
&&\theta_{\mu \mu}=\left[\frac{\beta(g)}{2g}\right]F^a_{\mu \nu}F^{a \mu \nu}
\nonumber\\
&&\partial^{\mu}J_{\mu 5}=-\left[\frac{\beta(g)}{2g}\right]F^a_{\mu\nu}\tilde{F}^{a\mu\nu}
\nonumber\\
&&\partial^{\mu}(x_{\lambda}\gamma^{\lambda})S_{\mu}=\gamma^{\mu}S_{\mu}
=2\frac{\beta(g)}{2g}F^a_{\mu\nu}\sigma_{\mu\nu}\lambda^a.
\label{anom}
\end{eqnarray}

This model has been analyzed in the presence of a mass term for both cases when m is small \cite{Masiero} and where
m is large \cite{Schechter1} and gluinos decouple. We are interested mainly in the trace anomaly from the perspective developed in these two cases.
We start with the case $m\ll\Lambda$ where the symmetry breaking term is,
\begin{equation}
{\cal L}_{\rm break}=mS_{\theta=0}+h.c.
\label{sb}
\end{equation}
We will ignore in what follows the effect of a nonzero vacuum angle $\theta_{\nu}$. The symmetry breaking terms
induces a non zero vacuum expectation value,
\begin{equation}
\frac{g^2}{32\pi^2}F^2=-\frac{m}{N_c}\langle\Phi\rangle+h.c.
\label{first564}
\end{equation}

But the super Yang Mills Lagrangian is just like QCD with a fermion in the adjoint representation so it is easy to compute the trace of the energy momentum tensor:
\begin{eqnarray}
&&\theta^{\mu}_{\mu}=-b\frac{g^2}{32\pi^2}F_{\mu\nu}^aF^{\mu\nu,a}+m(\langle\Phi\rangle+h.c.)=
\nonumber\\
&&=\frac{g^2}{32\pi^2}[(-3N_c)-N_c]F^2=-4N_c\frac{g^2}{32\pi^2}F^2.
\label{find}
\end{eqnarray}

Note that this corresponds to an effective coefficient of the beta function $b=4N_c$.

The trace of momentum energy tensor in Sannino Schechter approach \cite{Schechter1}, where for small masses the parameters $\delta=1$ and $\gamma=1$, has the form:
\begin{eqnarray}
\theta^{\mu}_{\mu}=-4N_c\frac{g^2}{32\pi^2}F^2.
\label{newsc}
\end{eqnarray}

Recently it has been shown \cite{Shifman} that non-supersymmetric $SU(N_c)$ Yang Mills theories with one spinor in the adjoint representation are planar equivalent to $SU(N_c)$ Yang Mills theories. Based on that Sannino and Shifman modifiy in \cite{SS} the VY Lagrangian by adding $1/N$ corrections which break supersymmetry and lead to the correct chiral and trace anomalies for the corresponding orientifold theories. Note here that Eqs.(\ref{first564}) and (\ref{find}) are in perfect agreement with (54)and (55) in \cite{SS} with an appropriate rescaling of the gluino field. This shows that our results are generic for this type of theories.

The effective beta function that one reads from Eq. (\ref{newsc}) corresponds to the exactly $4N_c$ zero modes of the gluons.
But we introduced a small mass term so that the gluinos do not decouple completely. We infer that gluinos zero modes
decouple while the gluino and gluon positive modes still cancel each other. Indirectly from \cite{Schechter1}but quite straightforward from \cite{Masiero} one can see that the contribution of the fermion condensate to the trace anomaly is equal to the contribution of the fermion zero modes to the beta function.

In \cite{Jora} we showed that chiral symmetry breaking can be interpreted as a collective effect due to the gluon and
fermion zero modes. If so it is natural to consider that in first order the contribution of the fermion condensate represents exactly the fermion zero modes in the beta function. In what follows we assume that, quite generally for
a non abelian gauge theory supersymmetric or not, the fermion condensate contribution to the trace anomaly in first order is,

\begin{eqnarray}
\sum_fm_f\langle{\bar q}q\rangle=(N_{f})_0\langle \frac{\beta(g)}{2g}G^{a\mu\nu}G^a_{\mu\nu}\rangle.
\label{genform}
\end{eqnarray}

where $(N_f)_0$ is the contribution of fermion zero modes to the beta function.

We will give a formal proof and an estimate for the gluon condensate in the next section.

In \cite{Schechter1} the authors introduce a straightforward procedure for decoupling the
heavy gluinos while matching the beta function at the scale of gluino masses. There it was found that a symmetry breaking term of the form $-m^{\delta}\Phi^{\gamma}$ where
\begin{eqnarray}
&&\gamma=\frac{12}{11}
\nonumber\\
&&\delta=4-3\gamma
\label{num}
\end{eqnarray}

satisfies all the requirements.
If we separate the degrees of freedom in both supersymmetric Yang Mills and Yang Mills we observe that in beta function for
supersymmetric QCD only the zero modes contribute as follows: $4N$ for the gluons and $N$ for gluinos (Actually the number of gluino modes is 2N but for us relevant is only the full contribution to the beta function). For Yang Mills
the zero modes gluon contribution is $4N$ whereas the positive mode gluon and ghost contribution is $-\frac{N}{3}$.
Note that for simplicity we do not separate in the case of positive modes the gluon and ghost contribution.
As mentioned in the first section for small gluino mass the beta function evolves according to a picture where
zero gluino modes decouple whereas positive modes are still canceled by those of the gluons. We would like
to expand the full potential introduced by Sannino and Schechter in terms of positive modes contribution.
This corresponds to an expansion in $\frac{1}{3}$ (Please note that QCD's expansion in $\frac{1}{N_c}$ is also an expansion in $\frac{1}{3}$).

We obtain:
\begin{eqnarray}
V_{SB}=-m^{\delta}\Phi^{\gamma}&=&
m\Phi(1+\frac{|N_{+}|}{|N_0|}\ln{\frac{\Phi}{m^3}}+...)=
\nonumber\\
&=&=m\Phi(1+\frac{1}{12}\ln{\frac{\Phi}{m^3}}+...).
\label{exp3}
\end{eqnarray}

Thus it is possible to see explicitly how the positive modes intervene as $m$ increases. It can be checked easily
that the approximate potential leads to the corresponding decoupling of the gluon field.

\section{Super QCD}
The super QCD Lagrangian was derived such that to satisfy all SUSY QCD anomalies  by Taylor, Veneziano and Yankielowicz in \cite{Veneziano2}:

\begin{eqnarray}
W_{TVY}=S
\left[
\ln (\frac{S^{N_c-N_f}\det{T}}{\Lambda^{3N_c-N_f}})-(N_c-N_f)\right].
\label{yank78}
\end{eqnarray}
From this one can obtain the potential:
\begin{eqnarray}
V_0(F,\Phi,F_T,t)=
-F\ln (\frac{\Phi^{N_c-N_f}\det{t}}{\Lambda^{3N_c-N_f}})-\Phi {\rm Tr}[F_Tt^{-1}]+ h.c.
\label{potsqcd}
\end{eqnarray}

In \cite{Schechter1} the authors use for simplicity a clever composite operator:
\begin{eqnarray}
Y^{N_f}=\frac{\Phi^{N_f}\det{F_T}}{\det(t)}.
\label{op768}
\end{eqnarray}

 Then by equating  the beta function above scale m with the supersymmetric one and below m with that of QCD and adding a consistency condition the authors obtain a rather complicated dependence for the parameters $\delta$, $\gamma$,
 $\tilde{\delta}$ and $\tilde{\gamma}$:
 \begin{eqnarray}
 \gamma&=&\frac{12N_c-4N_f}{11N_c-2N_f}
 \nonumber\\
 \tilde{\gamma}&=&12\frac{3N_c-N_f}{38N_c-11N_f}.
 \label{joes}
 \end{eqnarray}

 These values are too complicated to get a meaningful expansion in terms of the zero and positive modes but the whole set-up in \cite{Schechter1} is extremely useful for another purpose. Let us add to Eq (3.13) in \cite{Schechter1} a small mass term for the quarks in the form $m{\rm Tr}(F_T)$. Then the trace anomaly after decoupling will read,
 \begin{eqnarray}
 \theta_m^m&=&4V-\left[4F\frac{\partial V}{\partial F}+3\Phi\frac{\partial V}{\partial \Phi}+4Y\frac{\partial Y}{\partial Y}+3F_T\frac{\partial V}{\partial F_T}+h.c.\right]
 \nonumber\\
 \theta_m^m&=&\frac{11}{3}N_c-\frac{2}{3}N_f+m{\rm Tr}F_T
 \label{tr657}
 \end{eqnarray}

 It is clear that this equals exactly the trace anomaly for QCD for the case with small quark masses. We will then eliminate $F_T$ from the equation of motion. Actually this is most likely a procedure for finding vev's of the fields since we do not involve the kinetic term but this is exactly our purpose. The new equation is:
 \begin{eqnarray}
 \frac{\partial V}{\partial (F_T)_{ij}}=-\frac{F}{\det F_T}\frac{\partial \det F_T}{(F_T)_{ij}}=0
 \nonumber\\
 \frac{\partial V}{\partial (F_T)_{ii}}=-\frac{F}{\det F_T}\frac{\partial \det F_T}{(F_T)_{ii}}+m=0
 \label{setc756}
 \end{eqnarray}
 We multiply first equation in(\ref{setc756}) by $(F_T)_{ij}$ and the second by $(F_T)_{ii}$ and add the two of them to get:
 \begin{eqnarray}
 -N_fF+m{\rm Tr}F_T=0
 \label{tr5467}
 \end{eqnarray}

Note that this result is quite general; we could make masses different with the same l.h.s. We conclude that according to the picture sketched in the previous section and consistent with what we propose in \cite{Jora} as the quarks condense they decouple from the energy momentum tensor and the contribution of the condensate is exactly that of fermion zero modes to the beta function. Of course this leads to a first order determination(from first principle) of the gluon condensate in terms of the quark one.

Thus for QCD with two light flavors we obtain
$\langle \alpha_s F^2\rangle\approx0.00003 \,GeV^4$. For QCD with three flavors with degenerate masses equal to the mass
of the s quark we obtain $\langle \alpha_s F^2 \rangle \approx (0.051-0.061)\, GeV^4$ whereas for QCD with three flavors each with its own mass we get $\langle  \alpha_s F^2 \rangle \approx 0.0177\, GeV^4$. Compare these with the phenomenological value of the gluon condensate $\langle \alpha_s F^2 \rangle =(0.07\pm0.01)\,GeV^4$ \cite{Narison} derived using hadron data and heavy quark mass splitting. We do not take into account in our derivation the existence of other flavors than the light ones.

We return to the case of decoupling gluinos and squarks from the supersymmetric potential.
We will not change anything regarding the gluino but use instead a standard mass term for the squarks of the form
$-\rho^a[{\rm Tr}(t^{\dagger}t)]^b$ and also keep all the terms in the Lagrangian. Then our potential will be,
\begin{eqnarray}
V_0(F;\Phi;F_T,t)=
-F\ln \left[\frac{\Phi^{N_c-N_f}\det{t}}{\Lambda^{3N_c-N_f}}\right]-
\Phi{\rm Tr}[F_Tt^{-1}]-m^{\delta}\Phi^{\gamma}-\rho^a[{\rm Tr}(t^{\dagger}t)]^b+h.c.
\label{newpot56}
\end{eqnarray}

and we have the following equation of motion for the fields $\Phi$ and $t_{ij}$:
\begin{eqnarray}
&&\frac{\partial V_0}{\partial \Phi}=-\frac{(N_c-N_f)F}{\Phi}-{\rm Tr}(F_Tt^{-1})
-\gamma m^{\delta}\Phi^{\gamma-1}
\nonumber\\
&&\frac{\partial V_0}{\partial t_{ij}}=
-\frac{F}{\det{t}}\tilde{C_{ij}}+\Phi {\rm Tr}[F_T t^{-1}\tilde{t}_{ij}t^{-1}]-b \rho^a
[{\rm Tr}(t^{\dagger}t)]^{b-1}t_{ij}^*.
\label{eqm456}
\end{eqnarray}

Here $\tilde {C_{ij}}$ is what is left from the minor corresponding to the element $t_{ij}$ and
$\tilde{t_{ij}}$ is the matrix where all elements are equal to zero except element (ij) which is equal to 1.
It is evident that it is impossible to solve the second equation in (\ref{eqm456}) but we do not aim to find the relevant QCD potential but barely to find solutions for $\delta$, $\gamma$,a and b which lead to consistent decoupling of $\Phi$ and t.
Let us write the trace of the energy momentum tensor according to the procedure outlined in \cite{Schechter1}
\begin{eqnarray}
\theta_m^m&=&4 V -[4F\frac{\partial V}{\partial F}+ 3\Phi\frac{\partial V}{\partial \Phi}+
+2t_{ij}\frac{\partial V}{\partial t_{ij}}+3[F_T\frac{\partial V}{\partial F_T}]=
\nonumber\\
&-&\Phi{\rm Tr}(F_Tt^{-1})-4m^{\delta}\Phi^{\gamma}-4\rho^a[{\rm Tr}(t^{\dagger}t)]^b
\label{tenen456}
\end{eqnarray}

Here we used the equations of motion to simplify the expression. The second term can be substituted easily from
Eq (\ref{eqm456}) while for the second we will multiply the relevant equation by $t_{ij}$ and sum over the indices i and j. Thus we obtain:
\begin{eqnarray}
m^{\delta}\Phi^{\gamma}&=&-\frac{(N_c-N_f)}{\gamma}F-\frac{1}{\gamma}\Phi{\rm Tr}[F_Tt^{-1}]
\nonumber\\
\rho^a[{\rm Tr}(t^{\dagger}t)]^b&=&\frac{-N_f}{b}+\frac{1}{b}\Phi{\rm Tr}[F_Tt^{-1}]
\label{fin435}
\end{eqnarray}

The trace of the energy momentum tensor is given by:
\begin{eqnarray}
\theta_m^m=\frac{4(N_c-N_f)}{\gamma}F+\frac{4}{\gamma}\Phi{\rm Tr}[F_Tt^{-1}]+\frac{4N_f}{b}-\frac{4}{b}\Phi{\rm Tr}[F_Tt^{-1}]-\Phi{\rm Tr}[F_Tt^{-1}]
\label{fin678}
\end{eqnarray}

Since this should not depend on anything else except F we immediately have the condition:
\begin{equation}
\frac{4}{\gamma}-\frac{4}{b}-1=0
\label{cond647}
\end{equation}
 Then Eq (\ref{cond647}) together with the condition that after decoupling $\theta_m^m$ should be given by,

 \begin{eqnarray}
 \theta_m^m=(\frac{11}{3}N_c-\frac{2}{3}N_f)(F+F^*)
 \label{cond5467}
 \end{eqnarray}

 completely determines $\delta$, $\gamma$, a and b as:
\begin{eqnarray}
&&\gamma=\frac{12N_c}{11N_c+N_f}
\nonumber\\
&&b=\frac{12N_c}{8N_c+N_f}
\nonumber\\
&&\delta=4-3\gamma
\nonumber\\
&&a=4-4b
\label{param978}
\end{eqnarray}

Before going further let us see that for small masses we achieve a picture where zero modes for gluinos decouple while positive modes for both gluino and squarks cancel each other (There are no zero squarks modes).
Thus in this case, $\gamma=1$ and $b=\frac{4}{3}$ and this leads to:
\begin{eqnarray}
\theta_m^m=4N_c-4N_f+N_f(4-1)=4N_c-N_f
\label{dec45}
\end{eqnarray}

consistent to a beta function that contains only zero modes of gluon and quarks. Then in order
to add small squarks masses to the lagrangian one would need a mass term of the type $\rho^{-4/3}[{\rm Tr}(t^{\dagger}t)]^{4/3}$ to consistently decouple the term $\Phi{\rm Tr}[F_Tt^{-1}]$.
In order to see how the positive modes intervene in the process of decoupling we will consider,
as previously an expansion in $\frac{1}{3}$ taken as a small parameter. Then,
\begin{eqnarray}
\rho^a[{\rm Tr}(t^{\dagger}t)]^b&=&
\rho^{-\frac{4}{3}}[{\rm Tr}(t^{\dagger}t)]^{\frac{4}{3}}
\left[ 1+\frac{4(N_c-N_f)}{8N_c+N_f}\ln[\frac{[{\rm Tr}(t^{\dagger}t)]}{\rho^4}]+...\right]
\nonumber\\
m^{\delta}\Phi^{\gamma}&=&
m\Phi[1+\frac{N_c-N_f}{11N_c+N_f}\ln[\frac{\Phi}{m^3}]+...].
\label{exp76859}
\end{eqnarray}

\section{Discussion}

It is well known that the beta function for super Yang Mills (super QCD) can be expressed entirely in terms
of the zero modes for gluons, gluinos and fermions. In the transition to ordinary Yang Mills or QCD gluinos and squarks decouple thus releasing the positive modes for gluons and fermions. In the present work we have shown using  the VY and TVY potentials that this process takes place in two main stages dependent on the magnitude of the mass terms for gluinos and squarks: in the first stage the zero modes for gluinos decouple whereas the positive modes for gluon and gluinos (quarks and squarks) still cancel each other. Then at a larger mass term the decoupling becomes complete.
This evolution can be traced by using two mass terms instead of the usual single one, each corresponding to a definite stage.

Moreover, in the same framework and quite generally, we obtain that the contribution of the fermion condensates (quarks and gluinos) to the trace of the energy momentum is equal in magnitude and opposite in sign with the contribution of fermion zero modes to the beta function. Thus a small mass term in the Lagrangian leads to the decoupling of fermion zero modes. It is then possible to estimate the gluon condensate for QCD with two or three quark flavors.
We assumed that the TVY potential works also for $N_f=N_c$ when baryon degrees of freedom may intervene.
The case $N_f=N_c$ will be analyzed thoroughly in another work.

\section*{Acknowledgments} \vskip -.5cm
We are happy to thank R. Escribano and M. Jamin for useful related discussion. We are grateful to A. Fariborz and J. Schechter for support and encouragement and J. Schechter for useful comments on the manuscript.
This work has been supported by CICYT-FEDER-FPA2008-01430.

\end{document}